# Non-Hermitian Ring Laser Gyroscopes with Enhanced Sagnac Sensitivity

Mohammad P. Hokmabadi[1], Alexander Schumer[1,2], Demetrios N. Christodoulides[1], Mercedeh Khajavikhan[1*]

[1]CREOL, College of Optics & Photonics, University of Central Florida, Orlando, Florida 32816, USA

[2]Institute for Theoretical Physics, Vienna University of Technology (TU Wien), 1040, Vienna, Austria, EU

[*]Corresponding author: mercedeh@creol.ucf.edu

**Gyroscopes play a crucial role in many and diverse applications associated with navigation, positioning, and inertial sensing [1]. In general, most optical gyroscopes rely on the Sagnac effect- a relativistically induced phase shift that scales linearly with the rotational velocity [2,3]. In ring laser gyroscopes (RLGs), this shift manifests itself as a resonance splitting in the emission spectrum that can be detected as a beat frequency [4]. The need for ever-more precise RLGs has fueled research activities towards devising new approaches aimed to boost the sensitivity beyond what is dictated by geometrical constraints. In this respect, attempts have been made in the past to use either dispersive or nonlinear effects [5-8]. Here, we experimentally demonstrate an altogether new route based on non-Hermitian singularities or exceptional points in order to enhance the Sagnac scale factor [9-13]. Our results not only can pave the way towards a new generation of ultrasensitive and compact ring laser gyroscopes, but they may also provide practical approaches for developing other classes of integrated sensors.**

Sensing involves the detection of the signature that a perturbing agent leaves on a system. In optics and many other fields, resonant sensors are intentionally made to be as lossless as possible so as to exhibit high quality factors [14-17]. As a result, their response is governed by standard perturbation theory, suited for loss-free or Hermitian arrangements [15]. Recently, however, there has been a growing realization that non-Hermitian systems biased at exceptional points (EPs) [18], can react much more drastically to external perturbations [8,9,19]. This EP-enhanced sensitivity- a direct byproduct of Puiseux generalized expansions- is fundamental by nature. In particular, for a system supporting an Nth order exceptional point, where N eigenvalues coalesce and their corresponding eigenvectors collapse on each other, the reaction to a perturbation ($\epsilon$) is expected to follow an Nth root behavior ($\epsilon^{1/N}$) [18]. This is in stark contrast to Hermitian systems where the sensing response is at best of order $\epsilon$. Given that $\epsilon^{1/N} \gg \epsilon$ for $|\epsilon| \ll 1$, this opens up entirely new possibilities for designing ultrasensitive sensors based on such non-Hermitian spectral singularities. For illustration purposes, Fig. 1 provides a comparison between the eigenvalue surfaces associated with a Hermitian (Fig. 1a) two level system ($N = 2$) and its corresponding non-Hermitian counterpart (Fig. 1b) when plotted in a two-parameter space around the spectral degeneracies. As shown in Fig. 1b, the presence of an exceptional point forces the two Riemann manifolds to become strongly intertwined with each other- an attribute that could in turn be used to augment a sensor's performance [20,21].

Given that sensing is universally important in many fields, the emerging idea of boosting the sensitivity of a particular system via non-Hermitian degeneracies could have substantial ramifications across several technical areas. Here, we show that the sensitivity of a standard Helium-Neon (He-Ne) ring laser gyroscope can be significantly enhanced provided that its resonator is judiciously modified so as to support an exceptional point. Figure 2 depicts a schematic of the non-Hermitian ring laser gyroscope, as used in this study. As opposed to a standard RLG, the retrofitted cavity involves a Faraday rotator along with a half waveplate (HWP). As we will see, these two elements acting in conjunction with the Brewster windows (BW), incorporated on both ends of the He-Ne gain tube, can now be used to introduce a differential loss contrast (or gain) $\Delta\gamma$, between the clockwise (CW) and the counterclockwise (CCW) lasing modes. The way this is achieved is depicted in Fig. 2a, where the evolution of the state of the polarization associated with the two counter-rotating modes is provided at three consecutive points (A, B, C) in the cavity. In this particular arrangement, the Brewster windows allow only x-polarized light to circulate in the cavity while rejecting the y-component. As a result, the CW mode enters the Faraday rotator (FR) as x-polarized at point A. Because of the magneto-optic effect, the polarization then subsequently rotates by a small angle $\alpha$ (point B). Under the action of the half waveplate the angle between the linear electric field component and the preferred x-axis is $2\beta - \alpha$ (point C), where the small angle $\beta$ denotes the orientation of the waveplate's principal axes with respect to the x-y coordinates. On the other hand, because of non-reciprocity, the CCW mode while it also starts as x-polarized at point C, it exits at an angle $2\beta + \alpha$ with respect to the x-axis (point A) after traversing these same two optical components. Therefore, as clearly indicated in Fig. 2a, the CW mode is expected to experience lower losses than its CCW counterpart does, after passing through the Brewster windows of the He-Ne tube. Hence, a differential loss ($\Delta\gamma$) can be introduced between these two counter-rotating modes. Finally, in order to establish an exceptional point in this cavity, it is then necessary to antagonize this differential loss with a mode-coupling process [20]. In our system, the coupling between the CW and CCW modes is readily induced using a weakly scattering object (SC), like for example an etalon, as shown in Fig. 2a. The aforementioned processes can be formally described by employing a Jones matrix approach for the elements involved (HWP, FR, BW, SC, etc), where the polarization state of the CW/CCW waves can be monitored after each pass through the following transfer matrix $\boldsymbol{T} = \boldsymbol{S_{SC}} \cdot \boldsymbol{P} \cdot \boldsymbol{J_{HWP}} \cdot \boldsymbol{J_{FR}} \cdot \boldsymbol{J_{BW}}$ (see methods). In this last expression, $\boldsymbol{S_{SC}}$ represents a conservative scattering matrix (producing coupling) while $\boldsymbol{P}$ denotes a phase accumulation matrix [22,23] that can in principle account for the Sagnac shift [3].

To experimentally demonstrate this enhanced Sagnac sensitivity, we use a custom-made, educational-grade He-Ne RLG (purchased from Luhs Company) [24]. The length of the triangular cavity is 138 cm, supporting a free spectral range of ~216 MHz at 632.8 nm. The maximum loss that can be afforded in this system is approximately ~3.6%. This resonator is then retrofitted with a Terbium Gallium Garnet (TGG) Faraday element that can provide up to ~4° rotation at a magnetic induction of ~80 $mT$. This is used in conjunction with a half waveplate whose rotation angle can vary in a controlled manner with a resolution of 0.005 degrees. An etalon in the cavity promotes lasing a specific longitudinal mode while providing some level of coupling between the CW/CCW modes. Other elements like the TGG also contribute to this coupling. Overall, the system is designed to allow maximum tunability in establishing an EP.

Figures 2b and 2c provide a comparison between the principles of operation behind a standard ring laser gyroscope and that of the EP-based RLG arrangement used in this study. In the first configuration, the Sagnac effect produces a shift ($\pm\Delta\omega_s/2$) in the lasing CW/CCW angular frequencies

(which at rest coincide at $\omega_0$), where the beating frequency $\Delta\omega_s = 8\pi A\Omega/\lambda_0 L$ depends on the angular velocity $\Omega$ of the rotating frame, the area $A$ enclosed by the light path having a perimeter L, as well as on the emission wavelength $\lambda_0 = 2\pi c/\omega_0$. Evidently, the beating frequency $\Delta\omega_s/2\pi$ in this Hermitian setup (that is electronically detected) is always proportional to $\Omega$ and is dictated by geometrical constraints (Fig. 2b). The situation is entirely different for the non-Hermitian configuration, where the carrier frequency $\omega_0$ can split by $\pm\Delta\omega_c/2$ even in the absence of the any rotation because of coupling effects arising from the scatterer (Fig. 2c). In turn, in this same static frame, by adjusting the differential loss $\Delta\gamma$, these two resonances can fuse with each other once again at $\sim\omega_0$, thus marking the presence of an exceptional point. After the system is set at an EP, upon rotation $\Omega$, the Sagnac shifts $\pm\Delta\omega_s/2$ now induce two new frequency lines at $\omega_0 \pm \Delta\omega_{EP}/2$ (Fig. 2c). In this case, the beating frequency $\Delta\omega_{EP}$ is no longer proportional to $\Omega$ but instead varies in an enhanced fashion since in this regime $\Delta\omega_{EP} \propto \sqrt{\Omega}$, as expected when operating at the vicinity of an exceptional point (Fig. 2c).

The frequency eigenvalues of the non-Hermitian RLG can be directly obtained from the transfer matrix **T** after imposing periodic boundary conditions. From here, the induced non-Hermitian splitting $\Delta\omega_{EP}$ can be obtained, which remains interestingly enough unaffected even in the presence of gain saturation (see Methods). Based on these results, under rest conditions, one can compute the frequency split associated with the CW/CCW counter-propagating modes in our actual system, as a function of the HWP angle, when for example the coupling strength is $\kappa = 400$ kHz (Fig. 3a). In this case, an exceptional point appears at $\beta \sim 4.7°$. The corresponding magnitude of the system's complex eigenvalues $|\Lambda_{1,2}|$ is plotted in Fig. 3b. The frequency beating signals expected from either the Hermitian (orange line) or non-Hermitian (black line) version of this same RLG are plotted in Fig. 3c as a function of $\Omega$. In the non-Hermitian case, we assume that the system was positioned at an EP ($\kappa = \Delta\gamma$) when again $\kappa = 400$ kHz. The EP-enhancement in the Sagnac shift is evident in this figure. For these parameters, if for example the system rotates at $\Omega = 1°/s$, the Sagnac signal from the unmodified version of this RLG (Hermitian) is approximately $\sim 7.325$ kHz while from the retrofitted one (EP-based) is expected to be $\sim 5.2$ times larger. Finally, Fig. 3d shows the beat note change as a function of gyration speed when the system deviates from the EP (by 0.05% to 0.1% of the coupling strength). While ideally one must keep the system at the exceptional point, for small deviations, the resulting error appears to be negligible.

Figure 4a displays a picture of the gyroscope setup where the detection of the beat note frequency is performed by externally interfering the clockwise and counterclockwise beams. Figure 4b depicts experimental results obtained from our RLG system when biased at an EP. In our experiments, prior to each set of measurements, the system was positioned at an exceptional point by monitoring the beat note as a function of the rotation angle of HWP (gain-loss contrast), i.e. setting the beat frequency as close as possible to zero. Data corresponding to three different coupling strengths are provided along with those from the standard unmodified arrangement. These results are plotted in a log-log scale as a function of the rotation rate $\Omega$ when $\kappa = 65, 150, 425$ kHz. While the response of the standard configuration is linear with respect to $\Omega$ (slope=1), the same is not true in its non-Hermitian embodiment. In the latter case, the response was found to vary in a square root manner, as evident from the slope of the accompanying three curves, which is very close to ½- a clear indication that an EP is at play. Our experimental observations clearly show that the scale factor of the Sagnac effect is significantly boosted by exploiting the very properties of EPs. The resulting Sagnac enhancement factors (with respect to the standard arrangement) are plotted in Fig. 4c for these same three cases. For $\kappa = 425$ kHz, more than an order of magnitude boost in sensitivity is observed when $\Omega = 0.4°/s$ (The reported minimum gyration speed, $\Omega = 0.1°/s$, is imposed by the limited rotation capability of the apparatus).

Several issues must be considered when using non-Hermitian arrangements for sensing purposes. First and foremost is appreciating the difference between sensitivity and detection limit [25]. In non-Hermitian settings, the sensitivity enhancement is a fundamental feature that is dictated by mathematical properties, governed by the perturbation expansion around an exceptional point. The detection limit on the other hand depends on the physical system and is primarily determined by the net gain (or loss), as well as the correlation between the laser noise associated with the two resonances [26,27]. In this regard, one in principle can increase the net gain, while keeping the RLG at the exceptional point by managing the gain contrast in order to boost both the sensitivity and detection limit- as we did in our design. Another technical issue is how closely one can reach and stabilize the system at the exceptional point [28,29]. In our current experiment, we fully rely on positioning the RLG at the EP prior to each set of measurements, i.e. by visually monitoring the beat note as a function of the rotation angle of HWP (gain-loss contrast). In devices to be used in the field, one may need to actively control the system to remain biased at the exceptional point. Such approaches have been suggested eleswhere [10,30].

In conclusion, we have demonstrated for the first time a new class of non-Hermitian ring laser gyroscopes that can display an enhanced Sagnac sensitivity. This is accomplished by exploiting the intriguing properties of a special family of non-Hermitian spectral singularities- the so-called exceptional points- where the RLG responds to the gyration speed in a square root manner, as opposed to being linear in standard arrangements. The proposed configuration may inspire new technological developments in various settings where measuring low rotation rates via ultra-compact systems is highly attractive. Finally, the very idea of transforming a standard measuring apparatus into an EP-based device with superior sensitivity may have important ramifications in other areas of science and technology.

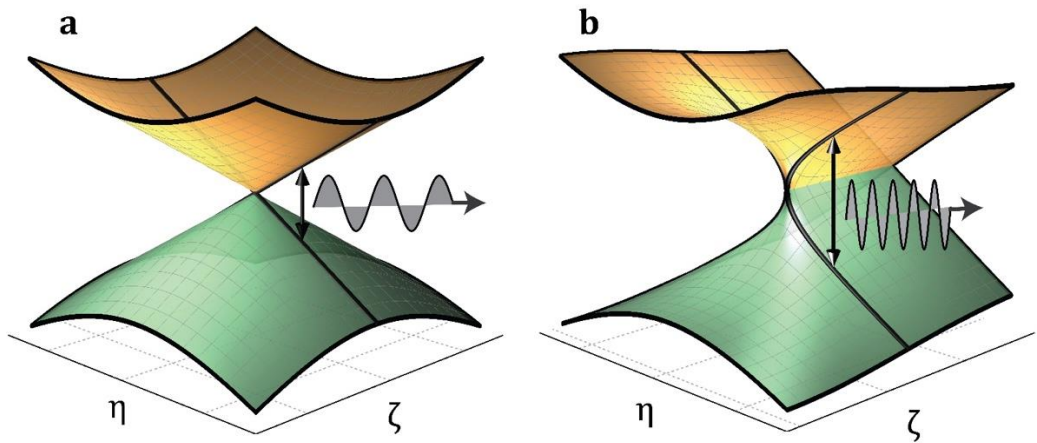

**Fig. 1 | Conceptual illustrations comparing the eigenvalue surfaces associated with Hermitian and non-Hermitian two-level systems. a.** The real part of the eigenvalues plotted in parameter ($\eta - \xi$) space (normalized detuning versus normalized gain-loss contrast) when the arrangement is Hermitian. Because of Hermiticity, this system responds linearly to perturbations. **b.** The real part of the eigenfrequency surface corresponding to a non-Hermitian configuration in the same parameter space. In the presence of an EP, the two Riemann manifolds are strongly intertwined, leading to a square root response to perturbations- as indicated by the frequency of the emitted signal. Using this system, an enhanced sensitivity to small changes is expected.

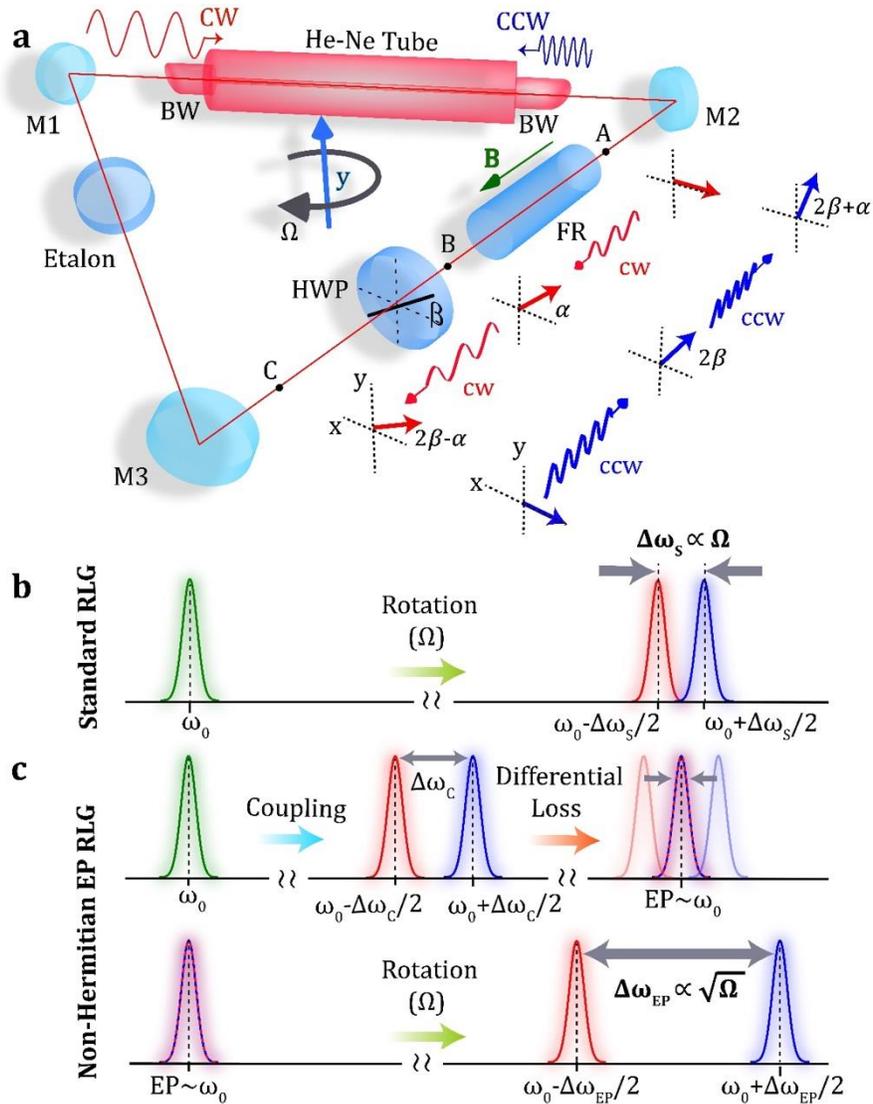

**Fig. 2 | Principle of operation of an EP-based He-Ne ring laser gyroscope. a.** The equilateral RLG cavity is comprised of three highly reflective mirrors ($M_{1,2,3}$), a He-Ne gas tube as the gain medium, and an etalon to select the desired longitudinal mode/s. In contrast to a standard RLG, the EP-based cavity also includes a Faraday rotator (FR) along with a half waveplate (HWP). These two elements, in conjunction with the Brewster windows (BWs), introduce a differential loss between the clockwise ($A \rightarrow C$) and counterclockwise ($C \rightarrow A$) directions. **b.** In a standard RLG, the Sagnac effect induces a shift ($\pm \Delta \omega_s/2$) in the stationary lasing angular frequencies ($\omega_0$) associated with the CW/CCW modes. The resulting beating frequency $\Delta \omega_s$ is proportional to the angular velocity $\Omega$ of the rotating frame. **c.** In an EP-based arrangement, the CW and CCW modes are first coupled to each other due to the presence of weak scattering in the system. Consequently, the stationary lasing frequency ($\omega_0$) splits according to the ensuing coupling strength ($\pm \Delta \omega_c/2$). On the other hand, the loss contrast ($\Delta \gamma$) induced by the simultaneous act of FR, HWP, and BWs brings the two split modes back to $\omega_0$, i.e. to an EP. Once the RLG is biased at the EP, the gyration will lead to a beat frequency which is now proportional to $\sqrt{\Omega}$. It is expected that for small rotation rates, the beat note of the EP-RLG will be considerably enhanced in comparison to that from a standard arrangement.

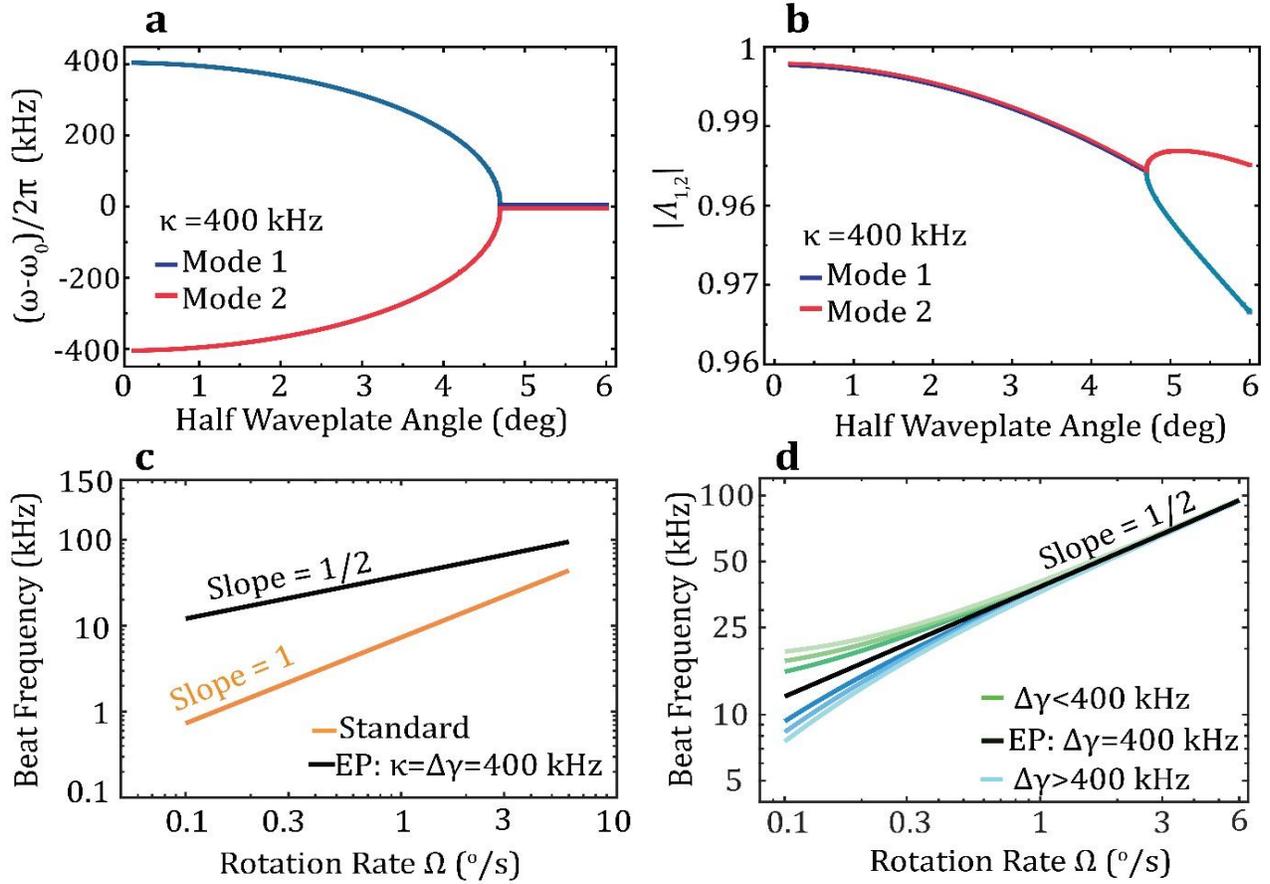

**Fig. 3 | Bifurcations of complex eigenfrequencies and sensitivity enhancement of EP-based RLG around an exceptional point a.** The resonance frequencies associated with the two coupled modes of an EP-based RLG at rest, versus the angle of HWP, when the coupling strength $\kappa$ is set to 400 kHz. These plots are obtained from Jones matrix analysis after considering gain saturation effects. The exceptional point in this system occurs at an HWP angle of ~4.7°. **b.** The magnitude of these same eigenvalues as a function of the HWP angle when the non-Hermitian RLG is stationary. **c.** The beat frequency as a function of the angular speed $\Omega$ (in log-log scale) for a standard RLG (orange) and a non-Hermitian RLG (black). The RLG is set exactly at the EP where the loss contrast balances the coupling ($\Delta\gamma = \kappa = 400$ kHz). For the standard RLG the slope of the curve is unity, while it is reduced to ½ for the non-Hermitian arrangement. **d.** The calculated beat frequency for the non-Hermitian RLG as a function of rotation rate $\Omega$, when the loss contrast does not exactly balance the coupling (the differential loss differs from the coupling by 0.05% to 0.1%) and hence the system is not precisely located at the EP. While the slope is approximately ½ at large rotation rates, it deviates from this value at small angular velocities when $\Delta\gamma \neq \kappa$. When $\Delta\gamma > \kappa$ (above EP) the beat frequency (blue lines) is below that of the ideal case (black line), which indicates a reduced sensitivity to rotation $\Omega$. On the other hand, when the system is biased below EP ($\Delta\gamma < \kappa$), the beat frequency does not exhibit a strong dependence on the gyration speed.

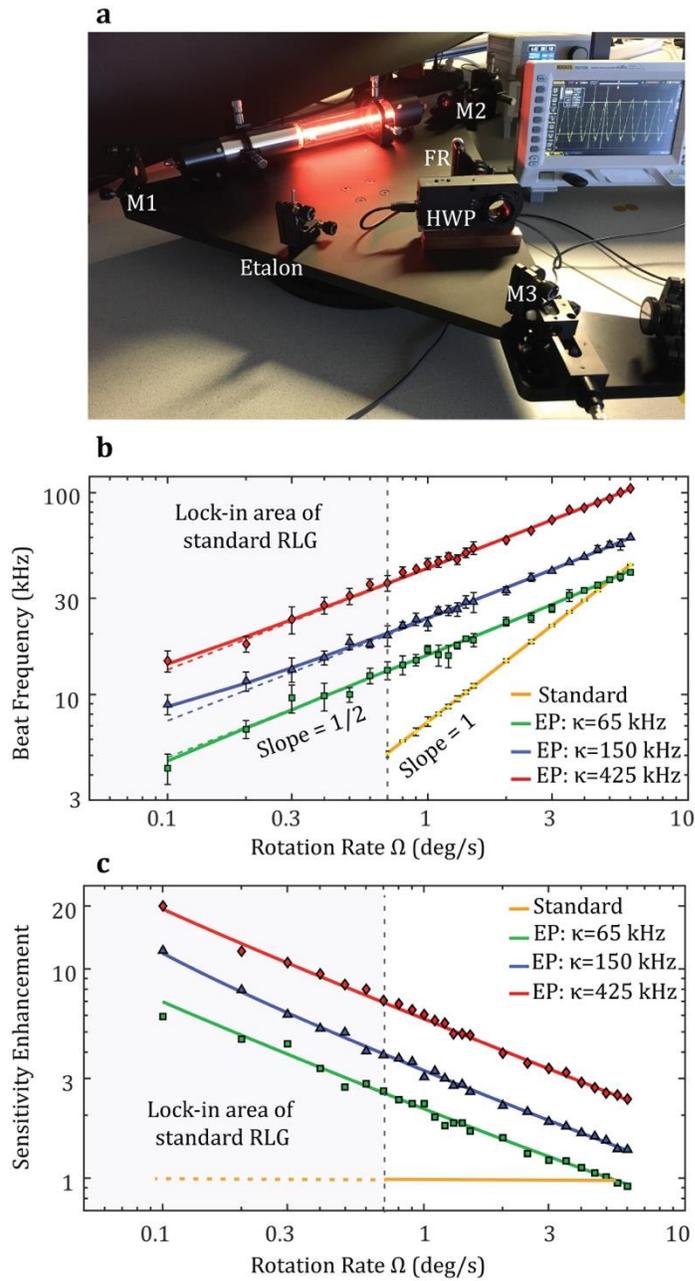

**Fig. 4 | Measured beat frequency and sensitivity enhancement factor versus rotation rate. a.** A picture of the RLG setup used in this study. **b.** Beat frequency versus rotation rate $\Omega$ (in log-log scale) for a standard RLG (yellow data marks) and a non-Hermitian RLG at three different coupling strengths, $\kappa_1 = 65$ kHz (green), $\kappa_2 = 150$ kHz (blue), and $\kappa_3 = 425$ kHz (red). The dashed lines correspond to theoretical calculations, where the non-Hermitian system is exactly biased at the EP ($\Delta\gamma_i = \kappa_i$). The solid lines represent fitted data when the system is slightly detuned from the EP (see Methods) (here $\Delta\gamma_1 = 1.0003\kappa_1, \Delta\gamma_2 = 0.9992\kappa_2, \Delta\gamma_3 = 0.9999\kappa_3$). The yellow line has a slope of unity indicating that the Sagnac shift in a standard cavity varies linearly with $\Omega$. Meanwhile, the slop associated with the non-Hermitian curves is approximately ½- indicating the presence of an EP. Moreover, whereas in the standard RLG, the lock-in effect limits gyration measurements below $\Omega = 0.7°/s$ (shaded region), the EP-based configuration is capable of detecting smaller rotation rates (only limited by the resolution of the step motor $0.1°/s$). **c.** Sensitivity Enhancement (S.E.), defined as the ratio of the non-Hermitian beat frequency to that of a standard RLG, is obtained from the measured data for the aforementioned three coupling strengths. For $\Omega < 0.7°/s$, the S.E. is calculated based on the anticipated value of the beat frequency from the standard RLG, provided that lock-in does not occur. The solid lines (red, blue, and green) represent theoretical curves corresponding to the parameters used in our experiments.

**Acknowledgements**

National Science Foundation (ECCS 1454531, DMR-1420620, ECCS 1757025), Office of Naval Research (N0001416-1- 2640, N00014-18-1-2347), Air Force Office of Scientific Research (FA9550- 14-1-0037), Army Research Office (W911NF-16-1-0013, W911NF-17-1- 0481), U.S.-Israel Binational Science Foundation (BSF) (2016381), DARPA (D18AP00058, HR00111820042, HR00111820038). The authors would also like to thank Dr. Walter Luhs for his help in setting up the gyroscope and for performing some of the initial measurements.

**Author contributions**

All authors contributed equally to this work.

**Competing interests**

The authors declare no competing interests.

**Additional information**

**Reprints and permissions information** is available at www.nature.com/reprints.
**Correspondence and requests for materials** should be addressed to mercedeh@creol.ucf.edu

# METHODS

**Analysis of exceptional points in a linear RLG cavity** The polarization state of the two counter-propagating CW/CCW beams can be described using Jones matrix formalism. In doing so, we first ignore any gain saturation effects. As we will see, while this approach is linear, it is still capable of predicting the correct lasing frequencies, which in this case happen to be identical to those expected from the nonlinear model. In order to simultaneously treat the CW and CCW fields around the cavity, we represent the polarization state as a $4 \times 1$ Jones vector $\left(E_{1,x}^{CCW}, E_{1,y}^{CCW}, E_{2,x}^{CW}, E_{2,y}^{CW}\right)^T$ where the indices 1 and 2 represent the two entry ports of each optical element involved. From this point on, the $4 \times 4$ Jones matrices can be set up. During a roundtrip, light in this RLG cavity is expected to first traverse the Brewster windows ($J_{BW}$), then the Faraday rotator ($J_{FR}$), the half-wave plate ($J_{HWP}$), the scattering element ($S_{SC}$) and also experience a phase accumulation ($P$) that could also account for the phase shifts $\pm \Delta\Phi$ induced by the Sagnac effect. If the cavity has the shape of an equilateral triangle, $\Delta\Phi = L^2 \pi \, \Omega / (3\sqrt{3} \, c \, \lambda)$. The Jones matrices associated with these elements are as follows:

$$S_{SC} = \begin{bmatrix} i\cos\sigma & 0 & \sin\sigma & 0 \\ 0 & i\cos\sigma & 0 & \sin\sigma \\ \sin\sigma & 0 & i\cos\sigma & 0 \\ 0 & \sin\sigma & 0 & i\cos\sigma \end{bmatrix}, \quad J_{BW} = \begin{bmatrix} 1 & 0 & 0 & 0 \\ 0 & 0 & 0 & 0 \\ 0 & 0 & 1 & 0 \\ 0 & 0 & 0 & 0 \end{bmatrix},$$

$$P = \begin{bmatrix} e^{i\left(\frac{\omega L}{c}+\Delta\Phi\right)} & 0 & 0 & 0 \\ 0 & e^{i\left(\frac{\omega L}{c}+\Delta\Phi\right)} & 0 & 0 \\ 0 & 0 & e^{i\left(\frac{\omega L}{c}-\Delta\Phi\right)} & 0 \\ 0 & 0 & 0 & e^{i\left(\frac{\omega L}{c}-\Delta\Phi\right)} \end{bmatrix},$$

$$J_{FR} = \begin{bmatrix} \cos\alpha & -\sin\alpha & 0 & 0 \\ \sin\alpha & \cos\alpha & 0 & 0 \\ 0 & 0 & \cos\alpha & \sin\alpha \\ 0 & 0 & -\sin\alpha & \cos\alpha \end{bmatrix},$$

$$J_{HWP} = \begin{bmatrix} \cos 2\beta & \sin 2\beta & 0 & 0 \\ \sin 2\beta & -\cos 2\beta & 0 & 0 \\ 0 & 0 & \cos 2\beta & \sin 2\beta \\ 0 & 0 & \sin 2\beta & -\cos 2\beta \end{bmatrix}. \quad (1)$$

Here, the rotation angle $\alpha$ of the Faraday rotator (FR) is fixed whereas the rotation angle $\beta$ of the half waveplate (HWP) can be adjusted to tune the differential loss in the system. The power reflectance of the scatterer is given by $R = \sin^2(\sigma)$. In general, the Brewster windows are expected to reject the $y$-polarization. After a roundtrip, the CW/CCW polarization components must repeat, therefore:

$$\left(E_{2,x}^{CCW}, E_{2,y}^{CCW}, E_{1,x}^{CW}, E_{1,y}^{CW}\right)^T = S_{SC} \cdot P \cdot J_{HWP} \cdot J_{FR} \cdot J_{BW} \cdot \left(E_{1,x}^{CCW}, E_{1,y}^{CCW}, E_{2,x}^{CW}, E_{2,y}^{CW}\right)^T, \quad (2)$$

where the roundtrip matrix $T = S_{SC}.P.J_{HWP}.J_{FR}.J_{BW}$ can provide the complex eigenfrequencies associated with the RLG cavity which are given by:

$$\Lambda_{1,2} = 0,$$

$$\Lambda_{3,4} = \frac{i}{2} e^{i\left(\frac{L\omega}{c} - \Delta\Phi\right)}\left[\cos(\sigma)\left(\cos(2\beta - \alpha) + e^{2i\Delta\Phi}\cos(2\beta + \alpha)\right) + Z\right] \quad (3)$$

$$Z = \pm\sqrt{\cos(\sigma)^2\left(\cos(2\beta - \alpha)^2 + e^{4i\Delta\Phi}\cos(2\beta + \alpha)^2\right) - (3 - \cos(2\sigma))e^{2i\Delta\Phi}(\cos(2\beta + \alpha)\cos(2\beta - \alpha))}$$

Two of the eigenvalues are zero because of the Brewster windows extinguishing the $y$-polarization. In this formalism, the EP is reached when the square root in equation (3) vanishes, which occurs at a half waveplate angle of

$$\beta_{EP} = \frac{\arctan(\cot(\alpha)\sin(\sigma))}{2}. \quad (4)$$

This parametrization for the critical angle $\beta_{EP}$ also emphasizes the necessity for a FR, as the $\cot(\alpha)$ is only defined when $\alpha \neq 0$, which implies that in the absence of a FR no EP can exist in this system. The eigenvalues of $T$ enable us now to calculate the lasing frequencies $\omega_i$, i.e. the frequencies for which the phase accumulation after one roundtrip equals a multiple integer of $2\pi$ : $\text{Arg}(\Lambda_i) = 2\pi m, m \in \mathbb{Z}$. The steady state lasing frequencies are hence found to be

$$\omega_i = \frac{c}{L}(2\pi m - \text{Arg}(\Lambda_i|_{L=0})). \quad (5)$$

Finally, the general beating frequency that stems from the splitting of the $m$-th mode turns out to be

$$\Delta\omega = \frac{c}{L}\text{Arg}\left(\frac{\Lambda_4}{\Lambda_3}\Big|_{L=0}\right), \quad (6)$$

where $\Lambda_{3,4}$ are given by equation (3).

**Gain saturation effects on the EP response** The RLG is assumed to lase in a single longitudinal mode $\omega_0$. Both the Sagnac effect and the scatterer are expected to induce a frequency splitting of a few hundred kHz, which is quite small compared to $\omega_0$. This in turn allows one to describe the dynamics of CW/CCW counter-propagating waves in terms of a coupled mode approach,

$$i\frac{dA(t)}{dt} + \omega_a A(t) + \kappa e^{i\phi} B(t) + i\gamma_a A(t) - \frac{ig_0 A(t)}{1+s(|A(t)|^2+|B(t)|^2)} = 0, \quad (7a)$$

$$i\frac{dB(t)}{dt} + \omega_b B(t) + \kappa e^{-i\phi} A(t) + i\gamma_b B(t) - \frac{ig_0 B(t)}{1+s(|A(t)|^2+|B(t)|^2)} = 0, \quad (7b)$$

where $A(t)$ denotes the complex amplitude of the CW while $B(t)$ represents that of the CCW. In the above equations, $\omega_{a,b} = \omega_0 \pm \delta$ are the detuned resonance frequencies, $\delta$ is a shift caused by Sagnac effect $\delta = \Delta\omega_s/2$, $\kappa$ is the coupling strength between the counter-propagating fields due to the scatterer, $\gamma_{a,b}$ are the linear loss rates, $g_0$ is the unsaturated gain and $s$ is a parameter that is inversely proportional to the saturation intensity. The detuning $\delta$ caused by the Sagnac effect in the equilateral cavity is given by $\delta = \Delta\omega_s/2 = L\pi\Omega/(3\sqrt{3}\lambda_0)$, where $\Omega$ is the angular velocity of the rotating frame, $L$ is the perimeter of the cavity and $\lambda_0 = 2\pi c/\omega_0$ is the stationary lasing wavelength. Now, by considering the normalized parameters of $\tilde{\delta} = \delta/\kappa$, $\tilde{\gamma}_{a,b} = \gamma_{a,b}/\kappa$, $\tilde{g}_0 = g_0/\kappa$ and $\tau = \kappa t$ as well as gauge transformations of $A(t) = e^{i(\omega_0 t+\phi/2)}a(t)/\sqrt{s}$ and $B(t) = e^{i(\omega_0 t-\phi/2)}b(t)/\sqrt{s}$, equations (7a) and (7b) can be represented in a dimensionless form shown below

$$i\frac{da(\tau)}{d\tau} - \tilde{\delta}\, a(\tau) + b(\tau) + i\tilde{\gamma}_a\, a(\tau) - \frac{i\tilde{g}_0 a(\tau)}{1+(|a(\tau)|^2+|b(\tau)|^2)} = 0, \quad (8a)$$

$$i\frac{db(\tau)}{d\tau} + \tilde{\delta}\, b(\tau) + a(\tau) + i\tilde{\gamma}_b\, b(\tau) - \frac{i\tilde{g}_0 b(\tau)}{1+(|a(\tau)|^2+|b(\tau)|^2)} = 0. \quad (8b)$$

The eigenmodes of the above system of equations can then be obtained by considering the modal amplitudes as $(a(\tau), b(\tau))^T = (a_0, b_0)^T e^{i\lambda\tau}$, $\lambda \in \mathbb{R}$ with $a_0, b_0$ as complex constants. Assuming that the eigenvectors of the above arrangement are expressed as $\vec{v} = a_0(1, \rho e^{i\theta})^T$, the relationship between $\rho$ and $\theta$ can be described as

$$\left(\rho - \frac{1}{\rho}\right)\cos(\theta) = 2\tilde{\delta}, \quad (9a)$$

$$\left(\rho + \frac{1}{\rho}\right)\sin(\theta) = 2\Delta\tilde{\gamma}, \quad (9b)$$

with $\Delta\tilde{\gamma} = \frac{\tilde{\gamma}_b - \tilde{\gamma}_a}{2}$ being the dimensionless differential loss. The structure of these two equations entails that it is sufficient to calculate a single solution pair $(\rho, \theta)$ to obtain all four possible real solutions: $(\rho, \theta), (\rho^{-1}, \pi - \theta), (-\rho, \pi + \theta)$ and $(-\rho^{-1}, -\theta)$.

The stationary system ($\tilde{\delta} = 0$) is found to be a PT-symmetric arrangement where the EP, located at $\Delta\tilde{\gamma} = 1$, separates the unbroken ($\Delta\tilde{\gamma} < 1$) from the broken PT-symmetry phase ($\Delta\tilde{\gamma} > 1$). In the former case the fields in the clockwise and counterclockwise directions have the same amplitude $|a_0|$ with $\rho = 1$ but are phase shifted by $\theta = \pm\arcsin(\Delta\tilde{\gamma})$. The eigenvalues are found to be $\lambda_{1,2} =$

$\pm \cos(\theta) = \pm\sqrt{1 - \Delta\tilde{\gamma}^2}$. In this regime, the resulting dimensionless frequency splitting due to coupling is expressed as $\Delta\tilde{\omega} = 2\sqrt{1 - \Delta\tilde{\gamma}^2}$ which exhibits the characteristic square root dependence in the vicinity of the EP. Once at the EP ($\Delta\tilde{\gamma} = 1$), the eigenvalues and eigenvectors coalesce ($\lambda_{1,2} = 0$, $\vec{v}_{1,2} = (1,\ i)^T$), rendering the eigenbasis defective. Even though the EP is a point of dimension zero, the remnants of this singularity manifest themselves as a square root dependence in the perturbation series (Puiseux series) of the eigenvalues and eigenvectors. To demonstrate this effect, the system is analyzed near the EP in the presence of a rotational perturbation ($\tilde{\delta} \neq 0$). Under such circumstances, it can be shown that equations (9a) and (9b) always have two distinct real solutions:

$$\rho = e^x, \quad \theta = \arcsin\left(\frac{\Delta\tilde{\gamma}}{\cosh(x)}\right), \quad (10)$$

where $\cosh(2x) = \Delta\tilde{\gamma}^2 + \tilde{\delta}^2 + \sqrt{4\tilde{\delta}^2 + (\Delta\tilde{\gamma}^2 + \tilde{\delta}^2 - 1)^2}$. Thus, the eigenfrequencies associated with the rotating RLG are given by

$$\lambda_{1,2} = \pm\sqrt{\cosh(x)^2 - \Delta\tilde{\gamma}^2} = \pm\sqrt{\frac{1 - \Delta\tilde{\gamma}^2 + \tilde{\delta}^2 + \sqrt{4\Delta\tilde{\gamma}^2\tilde{\delta}^2 + (1 - \Delta\tilde{\gamma}^2 + \tilde{\delta}^2)^2}}{2}}, \quad (11)$$

which can also be simplified as $\lambda_{1,2} = \Re\left(\pm\sqrt{1 + (\tilde{\delta} + i\,\Delta\tilde{\gamma})^2}\right)$ and thus, the dimensionless beat frequency is obtained as $\Delta\tilde{\omega} = 2|\lambda_1 - \lambda_2|$. Finally, from the imaginary parts of equations (8a) and (8b), we can derive the lasing condition which ensures that, even in the detuned case, both modes of the RLG oscillate

$$\tilde{g}_0 > \max\left(\frac{\tilde{\gamma}_a + \tilde{\gamma}_b}{2} - \Delta\tilde{\gamma}\tanh(x), \frac{\tilde{\gamma}_a + \tilde{\gamma}_b}{2} + \Delta\tilde{\gamma}\tanh(x)\right). \quad (12)$$

It becomes apparent that the unsaturated gain needs to be greater than the average loss, if the modes are not degenerate ($x \neq 0$).

**Linear versus non-linear eigenvalue analysis** The beat frequency from the linear and nonlinear cases can be compared, after defining the differential loss ($\Delta\tilde{\gamma}$) and the coupling strength ($\kappa$) used in equation (11). The latter can be obtained from equation (6) by using Jones matrix analysis, provided that the system is at rest ($\Delta\Phi = 0$) with no FR and HWP in action ($\alpha = \beta = 0°$). Under this condition, the coupling strength of the scatterer will be found as half of the beat frequency from equation (6) and is given by

$$\kappa = \frac{\Delta\omega}{2} = \frac{c\,\sigma}{L}. \quad (13)$$

On the other hand, the differential loss ($\Delta\tilde{\gamma} = \Delta\gamma/\kappa$) can be calculated by subtracting linear losses ($\gamma_{1,2}$) of CW/CCW modes within the cavity. A very accurate approximation for such losses ($\gamma_{1,2}$) can be derived assuming exponential damping of the intensity after each revolution in the RLG, i.e.

$$\frac{I_{n+1}^i}{I_n^i} = e^{-\frac{2\gamma_i L}{c}}, \quad i = \text{CW, CCW} \quad (14)$$

where $n$ is the number of roundtrips. Practically speaking, if one now ignores scattering losses and considers the effect of the BWs in extinguishing $y$-polarization as the only source of dissipation, then the differential loss $\Delta\gamma$ (in Hz) will only depend on the angle of the HWP and the FR

$$\Delta\gamma = \frac{c}{2L}\ln\left(\frac{\cos(2\beta-\alpha)}{\cos(2\beta+\alpha)}\right). \quad (15)$$

Replacing $\Delta\gamma$ and $\kappa$ from equations (13) and (15) in equation (11), will lead to a beat frequency from nonlinear solution ($\Delta\widetilde{\omega}_{NL}$) which will agree perfectly to that of the linear case ($\Delta\widetilde{\omega}_{JC}$), for all practical values $\alpha, \beta$, and $\kappa$

$$\Delta\widetilde{\omega}_{JC} = \frac{1}{\sigma}\text{Arg}\left(\frac{\Lambda_4}{\Lambda_3}|_{L=0}\right) \simeq 2\Re\left(\sqrt{1+\left(\tilde{\delta}+i\,\Delta\tilde{\gamma}\right)^2}\right) = \Delta\widetilde{\omega}_{NL} \quad (16)$$

The connection between the phase shift and the frequency detuning due to the Sagnac effect is given by $\delta = \frac{c}{L}\Delta\Phi$.

## Data availability

All data needed to evaluate the conclusions in the paper are present in the main text and methods. The datasets generated during and analyzed during this study are available from the corresponding author on reasonable request.